\documentclass[12pt,a4paper,english,superscriptaddress,aps,nofootinbib]{revtex4}
\usepackage[utf8]{inputenc}
\usepackage[T1]{fontenc}
\usepackage{ragged2e}
\usepackage{amsmath,amssymb,graphicx}
\makeatletter
\usepackage{babel}
\usepackage[active]{srcltx}
\usepackage{graphicx,color}
\usepackage{changebar}
\usepackage{hyperref}
\hypersetup{
    colorlinks=true,
    urlcolor= blue,
    citecolor=blue,
    linkcolor= blue}

\usepackage{amsmath}
\usepackage{braket}
\usepackage{dsfont}
\usepackage{mathtools}
\usepackage{slashed}
\usepackage{empheq}
\usepackage{tikz}
\usepackage{multirow}
\usepackage{subfigure}
\usetikzlibrary{decorations.pathmorphing}

\usepackage{graphicx}
\usepackage{amsfonts}
\usepackage{amsmath}
\begin{document}	

\title{Properties of a Majorana fermion ensemble with exciton-like mass}

\author{F. C. E. Lima}
\email{cleiton.estevao@fisica.ufc.br (F. C. E. Lima)}
\affiliation{Departamento de F\'{i}sica, Universidade Federal do Cear\'{a} (UFC), Campus do Pici, Fortaleza - CE, 60455-760 - Brazil}

\author{L. E. S. Machado}
\affiliation{Departamento de F\'{i}sica, Universidade Federal do Cear\'{a} (UFC), Campus do Pici, Fortaleza - CE, 60455-760 - Brazil}

\author{C. A. S. Almeida}
\email{carlos@fisica.ufc.br (C. A. S. Almeida)}
\affiliation{Departamento de F\'{i}sica, Universidade Federal do Cear\'{a} (UFC), Campus do Pici, Fortaleza - CE, 60455-760 - Brazil}

\begin{abstract}
\vspace{0.5cm}
\noindent \textbf{Abstract:} Considering the relativistic scenario, we dedicate our study to the relativistic quantum description of one-dimensional Majorana fermions. Thus, we focus on aspects related to exciton-like particles. Seeking to reach our purpose, one analyzes the relativistic quantum mechanical system characterized by an effective mass distribution. In this context, we adopt an exciton-like position-dependent mass without impurity, i.e., without electromagnetic interactions. From this perspective, one notes results of noteworthy interest as consequences of the theory adopted. For instance, we highlight that, even without interaction, exciton-like Majorana fermions manifest theoretically bound states. Also, we construct a Majorana fermion ensemble with effective mass immersed in a thermal reservoir. That allows for a thorough investigation of the thermodynamic properties of the system. Among the thermodynamic characteristics studied in the canonical ensemble, we focus on the Helmholtz free energy, mean energy, entropy, and heat capacity. The numerical results obtained for these thermodynamic properties corroborate the validity of the Dulong-Petit law for our system.\vspace{0.35cm}

\end{abstract}
	
\maketitle	

\section{Introduction}\justifying
    
    Within the microstructure of solids, a notable discrepancy in mass values is discernible between dynamic and static systems \cite{Ref1}. This differentiation leads us to the theory of effective mass \cite{Dresselhaus}. This concept has proven useful in describing electronic bands in the presence of perturbations, such as external magnetic fields and impurities in semiconductors \cite{Ref2}. Furthermore, one applies this theory to the quantum description of superlattices \cite{Ref3}, heterostructures \cite{Ref4}, and thermoelectric materials \cite{Ref5}.
    
    In a broad context, one reinterprets the effective mass as a mass distribution that varies with the position \cite{Lima1,Lima2}. In this scenario, the mass distribution delineates the spatial coordinates of charge carriers within the material's energy band, thereby engendering a system characterized by Position-Dependent Mass (PDM) \cite{Thomsen}. Since the inception of the PDM concept, this theoretical framework has demonstrated its pertinence by comprehensively addressing a spectrum of issues across diverse domains of physics \cite{Ref6,Ref7,Ref8}.
    
    The PDM formalism was employed by von Roos \cite{Ref9}, addressing non-homogeneous semiconductors. In pursuit of its objectives, von Roos utilizes the definition of the kinetic energy operator to describe the electrons in semiconductors. The exact form of this operator has been a subject of discussion due to the ambiguity of the so-called ordering parameters \cite{Thomsen,Ref10,Ref11,Ref12}. A particularly intriguing discussion involving the PDM concept emerges in the non-relativistic regime. In this scenario, there is an ambiguity related to the symmetrization of the kinetic energy operator \cite{Ref10,Ref11,Ref12,Gora}. Cavalcante et al. \cite{Ref13} propose the most suitable symmetrization in the non-relativistic limit using relativistic quantum mechanics. Subsequently, works adopting the PDM concept of some systems were considered to discuss solid-state physics problems \cite{Ref14}. Naturally, as reported in references  \cite{Ref13,Ref14}, in the context of relativistic quantum mechanics, the PDM concept makes it more appropriate to address solid-state physics problems since this ambiguity does not exist in this scenario.
    
    Adopting the PDM formalism in a relativistic quantum-mechanical context,  we explore the one-dimensional quantum-relativistic theory using the Majorana representation with PDM. The motivation for this investigation arises from the following inquiry: is it possible to describe an exciton-like system by considering the Majorana representation with PDM? To achieve our goal, we assume the fermionic description of Majorana \cite{Majorana}. In the Majorana representation, one describes the spinor from the theory as an entity such that $\Psi^c=\Psi$, i.e., the spinor is real \cite{Majorana,Elliot}. From this perspective, the equality between the spinor and its conjugate suggests the absence of a distinction between particle and antiparticle \cite{Majorana,Elliot}.
    
    The physical nature of Majorana's fermions has no experimental comprovation. However, it is possible to experimentally observe excitations concerning Majorana's condition in superconductors interface of the type wave-s \cite{Ref15} and strong-topological insulators \cite{Ref16}. Furthermore, there has been discussion about the potential of the so-called Majorana Zero Modes (MZM) applied to topological quantum computing \cite{Ref17}, characterizing them as a topic of growing interest in condensed matter physics.
    
    The Majorana condition is also applicable in semiconductor theory, where the study of charge carriers leads to the concept of excitons. Briefly, excitons are structures formed by the binding of an electron at the conduction band and a hole at the valence band by Coulomb's interaction \cite{Dresselhaus}. Thus, these excitations carry energy and momentum, presenting a null total charge. Considering that the exciton study requires effective mass approximation \cite{Ref18}, one can adopt the PDM concept in the relativistic theory. In this work, one studies a one-dimensional exciton-like fermionic system with PDM. We undertake this investigation to comprehend how the eigenstates of Majorana's fermions with PDM behave in the absence of impurities.
    
    This paper is structured as follows: we systematically examine the aspects of PDM formalism in Sec. II. Subsequently, in Sec. III, an analysis of the relativistic quantum system within Majorana's representation employing an exciton-like PDM is performed. One builds an ensemble of Majorana's particles with PDM at a thermal bath to study the thermodynamic properties of the system in Sec. IV. Finally, in Sec. V, we announce our discoveries.
    
    \section{On the effective mass}\justifying
    
    The theory of effective mass emerges in the description of graded mixed semiconductors \cite{Vliet}. In the derivation of this theory, one presupposes that the variation in chemical composition occurs slowly enough for configuration changes to become appreciable on extensive scales compared to the lattice constant \cite{Ref9,Ref18,Vliet,Slater}. Gora \cite{Gora}, Williams \cite{Gora}, van Vliet \cite{Vliet}, and Marshak \cite{Vliet} propose a theory based on Schr\"{o}dinger's equation of a compound semiconductor in Wannier's representation \cite{Ref18}. Following these discussions, one can demonstrate that the Hamiltonian of a PDM has various ways in the non-relativistic regime \cite{Ref9,Ref10,Ref11,Ref12,Gora}. However, Morrow \cite{Morrow} highlighted that the non-singularity of the Hamiltonians proposed by Gora and Williams \cite{Gora} and van Vliet and Marshak \cite{Vliet} persists when the wavelengths of the envelope functions are not short compared to the distance over which the chemical composition changes appreciably. Nevertheless, this precisely occurs in the case of excited surface impurity state wavefunctions that extend over hundreds of lattice sites.
    
    Using Bargmann's theorem \cite{Bargmann}, von Roos derives a unified Hamiltonian for PDM problems \cite{Ref9}. However, it was only in 1997 that Cavalcante et al. \cite{Ref13} a priori resolved the ambiguity in the non-relativistic Hamiltonian. In Ref. \cite{Ref13}, a relativistic approach is adopted, seeking the non-relativistic limit for the theory. In this case, the non-relativistic Hamiltonian for the system with PDM assumes the form proposed by Li and Kuhn \cite{Ref12}. Since these discussions, several works have emerged employing the theories with effective mass. For instance, one uses the PDM concept to investigate the quantum information of a solitonic mass distribution \cite{Lima1}. Meanwhile, one can use the PDM in explorations of non-Hermitian theories \cite{Lima2}. Thus, motivated by these applications, let us commence our investigation by studying the equation of motion for an electron in the energy band of a semiconductor with a wave vector $k$ associated with the group velocity
    \begin{align}\label{Eq1}
        v_{g}=\dfrac{1}{\hbar}\dfrac{dE}{dk},
    \end{align}
    where $E$ corresponds to the system's dispersion energy. Considering the action of an electric field $\epsilon$ on this electron over a time interval $\delta t$, one obtains the work $\delta E$, i.e.,
    \begin{align}\label{Eq2}
        \delta E=-e\epsilon v_{g}\delta t.
    \end{align}
    which leads us to
    \begin{align}\label{Eq3}
        \delta E=\hbar v_{g}\delta k.
    \end{align}
    
    Comparing the Eqs. \eqref{Eq1} and \eqref{Eq2}, we arrive at
    \begin{align}
    \label{Eq4}
        \delta k = -e\dfrac{\epsilon}{\hbar}\delta t,
    \end{align}
    and
    \begin{equation}\label{Eq5}
        \hbar\dfrac{dk}{dt}=-e\epsilon.
    \end{equation}
    
    One can write the Eq. (\ref{Eq5}) in terms of the external force, namely,
    \begin{align}\label{Eq6}
        F = \hbar\dfrac{dk}{dt}.
    \end{align}
    The Eq. (\ref{Eq6}) represents a crucial relationship determining the external force exerted on the electron within a crystal. The correlation between energy and wave vector manifests through the expression $E=\hbar^{2}k^{2}/2m$, where $\hbar^2/2m$ defines the curvature of the energy dispersion $E(k)$.
    
    We employ this approach in describing hole-electron pairs that constitute excitons in the energy bands. In this case, the holes possess the same expressions as the electrons for wave vector and energy but with opposite signs. Thus, differentiating the Eq. (\ref{Eq1}) concerning time, one obtains 
    \begin{align}\label{Eq7}
          \dfrac{dv_{g}}{dt}=\dfrac{1}{\hbar^{2}}\dfrac{d^{2}E}{dk^{2}}F.
    \end{align}
    
    Comparing the Eq. (\ref{Eq7}) with the definition coming from Newton's second law, we conclude that the effective mass is
    \begin{align}\label{Eq8}
       m(k) = \hbar^{2}\bigg(\dfrac{d^{2}E}{dk^{2}}\bigg)^{-1}.
    \end{align}
    
    \subsection{The effective mass for the one-dimensional exciton-like coupling}\justifying
    
    Using the definition $m(k)$, we will adopt the dispersion energy $E(k)$ considering the tight-binding approximation or strong binding \cite{Foulkes}. The tight-binding approximation follows the following premises \cite{Dresselhaus}: 
    \begin{enumerate}
        \item The eigenvalues and eigenfunctions of energy are known for an electron in an isolated atom;
        \item The atoms of the material remain spaced apart such that each electron finds a specific atomic site;
        \item One can approximate the periodic potential from the lattice by a superposition of interactions.
    \end{enumerate}

    \subsubsection{The linear combination of atomic orbitals}\justifying

By assuming that, at the neighborhood of each point of the lattice, the total Hamiltonian H approximates from the Hamiltonian of the localized atoms $H_a$, once bound states of the Hamiltonian $\phi(r)$ are well localized, we can express the relation
\begin{align}\label{RRev1}
    H_a\phi_j(r)=\varepsilon_j\phi_j(r),
\end{align}
where $\varepsilon_j$ are the energy eigenvalues and $\phi_j(r)$ can be negligible when $\vert r\vert$ reaches a distance on the order of the lattice parameter. This approximation becomes accurate by implementing the additional interaction term $\Delta U(r)$ in the Hamiltonian concerning the localized atoms, i.e.,
\begin{align}\label{RRev2}
    H=H_a+\Delta U(r).
\end{align}
Thus, for an atom at the origin, $\phi_j(r)$ is a good approximation of the stationary wave function for the total Hamiltonian. Similarly, the wave functions $\phi(r-R_n)$ is a good approximation for the corresponding lattice sites $R_n$ in the Bravais lattice. Therefore, if $\phi_j(r)$ satisfies Schrödinger's equation for localized atoms, as given by Eq. \eqref{RRev2}, one obtains that Schrödinger's equation for the total Hamiltonian given by Eq. \eqref{RRev1}, whenever $\Delta U(r) = 0$ and $\phi_j \neq 0$.

By considering that each atomic level $\phi_j(r)$ is $N$-th periodic level with wavefunctions $\phi_j(r - R_n)$, it is necessary to find the $N$ linear combinations of these states that represent the Bloch function $\phi_j(r, k)$ associated with the crystal's electrons. Based on the third premise mentioned above, we define the Bloch function in terms of the atomic orbitals as
\begin{align}\label{RRev3}
    \phi_j(r,k)=\frac{1}{\sqrt{N}}\sum_{R}\text{e}^{ik\cdot R}\phi(r-R), \hspace{0.25cm} j=1,2,\dots,n,
\end{align}
where $n$ is the wavefunction number in the unit cell. Thereby, we have $n$ wavefunctions for each wavevector $k$. Therefore, the eigenfunction for an electron $\psi_j(r,k)$ is
\begin{align}\label{RRev4}
    \psi_j(r,k)=\sum_{j'}C_{jj'}\phi_{j'}(r,k),
\end{align}
with $C_{jj'}$ being coefficients to be determined. The last two equations above lead us to wavefunction regarding atomic orbitals $\phi_j(r - R_n)$.

Meanwhile, the energy eigenvalues regarding the transfer and overlap matrices are
\begin{align}
    H_{jj'}=\langle\phi_j\vert H\vert\phi_{j'}\rangle \hspace{0.25cm} \text{and}
\hspace{0.25cm} S_{jj'}=\langle\phi_j\vert\phi_{j'}\rangle.
\end{align}
In this case, we have $j,j'=1,2,\dots,n$.

By substituting the Bloch function \eqref{RRev3}, one obtains 
\begin{align}
    H_{jj'}=\frac{1}{N}\sum_{R,R'}\text{e}^{ik(R'-R)}t^{jj'}_{RR'},
\end{align}
which defines the hopping parameter as
\begin{align}
    t^{jj'}_{RR'}=\langle\phi_j(r-R)\vert H\vert\phi_{j'}(r-R')\rangle.
\end{align}
In other words, this is the energy necessary for an electron to hop from one site at $R'$ to the position $R$. This parameter is typically negative and decreases as the distance between the sites increases \cite{Marder}. Similarly, one defines
\begin{align}
    S_{jj'}=\frac{1}{N}\sum_{R,R'}\text{e}^{ik(R'-R)}s^{jj'}_{R,R'},
\end{align}
i.e., the overlap parameter is
\begin{align}
    s^{jj'}_{R,R'}=\langle\phi_j(r-R)\vert\phi_{j'}(r-R')\rangle.
\end{align}

\subsubsection{The dispersion energy for the exciton}\justifying

In the absence of external factors such as lattice vibrations or impurities (electromagnetic fields), excitons are found in their free state, being able to interact with phonons and undergo the recombination process. In this configuration, the low binding energy of this arrangement results in a very short lifetime for these excitations, making the treatment of these structures challenging. On the other hand, the bound states can create energy levels that trap the excitons, reducing the recombination process and making them more stable. In this regard, despite the bosonic nature of the exciton, one can apply properties associated with it. For instance, its effective mass is specifically related to the dispersion energy, allowing for an analogous approach in fermionic systems, where the emergence of bound states is notorious.

Seeking to derive the dispersion energy, we consider the discussion presented in the previous subsection. Thus, let us assume that the Coulomb interaction term in the total Hamiltonian that characterizes excitons is zero since Majorana fermions have no charge. This assumption leads us to the unit cell, where each atom possesses only one orbital, $\phi(r)$, i.e.,
\begin{align}
    \psi_k(r)=\frac{1}{\sqrt{N}}\sum_m \text{e}^{ik\cdot R_m}\phi(r-R_m),
\end{align}
where the expected value of the Hamiltonian is
\begin{align}\label{RRRRRR12}
    \langle k\vert H\vert k\rangle=\frac{1}{N}\sum_m\sum_n\text{e}^{ik\cdot(R_n-R_m)}\langle\phi_m\vert H\vert\phi_n\rangle,
\end{align}
with $\phi_m=\phi(r-R_m)$. Therefore, if the particles are at the same site of the crystal lattice or among the first neighbors, one concludes that 
\begin{align}\label{RRRR0}
    \langle\phi_n\vert H\vert\phi_n\rangle=-\alpha \hspace{0.25cm} \text{and} \hspace{0.25cm} \langle\phi_m\vert H\vert\phi_n\rangle=-\gamma.
\end{align}
In this conjecture, note that $\alpha$ is the energy of an electron without the interaction with neighboring sites, and $\gamma$ is the hopping parameter concerning the probability of the electron moving to the first neighbors.

Outside the cases of Eq. \eqref{RRRR0}, one has $\langle\phi_m\vert H\vert \phi_n\rangle = 0$. Therefore, the relation \eqref{RRRRRR12} boils down to
\begin{align}
    E_k=\langle k\vert H\vert k\rangle=-\alpha-\gamma\sum_n \text{e}^{ik\dot R_n},
\end{align}
which leads us to
    \begin{align}\label{Eq11}
        E(k)=-\alpha-2\gamma \cos{ka}.
    \end{align}
    
\begin{figure}[!ht]
        \centering
    \subfigure[PDM at the reciprocal space.]{\includegraphics[height=6cm,width=7cm]{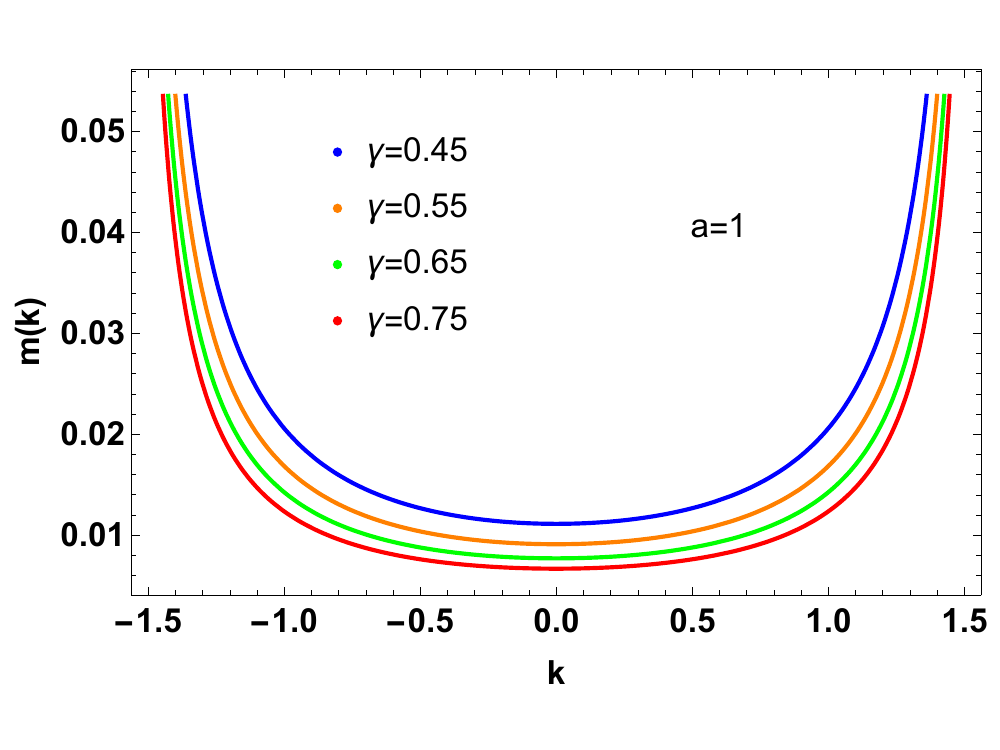}}
    \subfigure[PDM at the position space.]{\includegraphics[height=6cm,width=7cm]{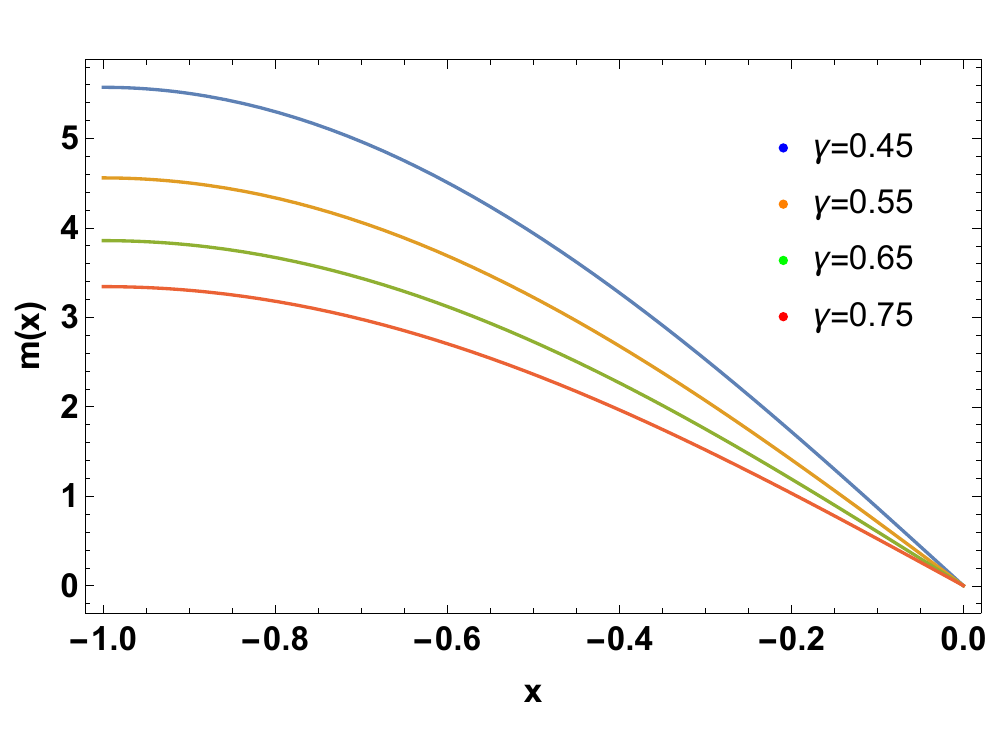}}\hfill
    \caption{Behavior of the PDM using several values of $\gamma$.}
        \label{fig1}
    \end{figure}

Therefore, the effective mass for this dispersion energy is
    \begin{align}\label{Eq12}
        m(k) = \dfrac{\hbar^{2}}{2\gamma a^2}\sec(ka).
    \end{align}
We present the behavior of the effective mass $m(k)$ in Fig. \ref{fig1}(a).
    
Note that the mass is at the reciprocal space $k$. Nevertheless, we wish to obtain a corresponding PDM. So, to reach our purpose, Fourier's transform is employed, considering the range corresponding to the first Brillouin zone\footnote{We consider the first Brillouin zone because the essential information on the energy and momentum of the crystal are in this domain.} $(-\pi/a \leq k \leq \pi/a)$. Thus, one obtains the exciton-like PDM, i.e.,
    \begin{align}\label{Eq13}
      m(x)=\frac{1}{2\gamma a^2\sqrt{2\pi}}\int_{-\pi/a}^{\pi/a}\text{sec}(k a)\cdot \text{e}^{-ikx}dk.
    \end{align}
One can obtain the result of the mass integration (\ref{Eq13}) using the residue theorem \footnote{The residue theorem is a method for calculating integrals used in analytical functions along closed paths that generalizes Cauchy's formula \cite{Nikiforov}. We expose further information regarding this calculation in Appendix A.}. In this case, the residue theorem leads us to
    \begin{align}\label{Eq14}
        m(x)= -\dfrac{\sqrt{2\pi}}{\gamma}\sin{\left(\dfrac{\pi x}{2}\right)}.
    \end{align}
For this calculation, we have assumed $a=1$. To obtain a positively defined PDM to the first Brillouin zone, one considers $x \in [-1, 0]$. We display the behavior correponding PDM in Fig. \ref{fig1}(b).

\section{Majorana's fermions with exciton-like PDM}\justifying
    
Let us focus our study on a specific class of particles, namely Majorana's fermions. Investigations and manipulations of Majorana's fermions in condensed matter systems are currently a topic of significant theoretical and experimental interest \cite{Alicea1,Alicea2}. As reported by Alicea \cite{Alicea1,Alicea2}, Majorana's fermions constitute ``half'' of an usual fermion. To create a fermion $f$, one requests the superposition of two Majorana's modes $\gamma_{1,2}$, where these modes are arbitrarily separated \cite{Alicea2}. Thus, Majorana fermions admit, in principle, the formation of bound states. That leads us to the following question: in a model describing a fermionic theory with an exciton-like PDM, in the absence of external confining fields (i.e., impurities), does the presence of bound states remain guaranteed? We address this question by adopting a fermionic formulation based on the one-dimensional Majorana representation with an exciton-like PDM. In this framework, we write
\begin{align}\label{Eq15}
    [\gamma^{\mu}(p_{\mu}-eA_{\mu})-m(x)]\psi(x,t)=0 \hspace{0.5cm} \text{with} \hspace{0.5cm} \mu=0,1;
\end{align}
where $m(x)$ is the PDM, $p_{\mu}$ is the particle's momentum, and $A_{\mu}=[V(x), 0]$ is the gauge field with $V(x)$ corresponding to the electric potential. One highlights that by adopting $A_{\mu}=[V(x), 0]$, only electric interactions are considered. Furthermore, the metric of the system is
    \begin{align}\label{Eq16}
        g_{\mu\nu}=\begin{pmatrix}
            +1 & 0\\
            0 & -1
        \end{pmatrix}. 
    \end{align}

The Majorana excitons are quasiparticles derived from conceptual extensions of the interaction between electron-hole pairs in semiconductor materials \cite{Kitaev}. In this framework, one notes that an electron-hole pair is a bosonic quasiparticle \cite{Loudon}. Naturally, these effective theories arise to describe the fermions that constitute the exciton, using analogies with these particles, as discussed in Refs. \cite{Alicea1,Alicea2}. Furthermore, Majorana  Hamiltonians contain topological symmetries described by some approaches \cite{Sato}. Thus, we can address these quasiparticles by adopting the creation and annihilation operator approaches, i.e., analogous relativistic Dirac descriptions \cite{Alicea1,Alicea2}. Another approach to treating these systems consists of considering the dispersion energy of these particles \cite{FLima11}. In this approach, the quasiparticle’s dispersion energy concerning a position-dependent mass allows us to treat the system as a PDM problem. In this case, we announced the Majorana excitons as an analogous effective Dirac theory with PDM. For the present study, we will address a Dirac PDM problem\footnote{In this work, we assumed the natural unit, i.e., $c = \hbar = k_B = 1$. In this framework, the energy has a GeV dimension.}, viz.,
    \begin{align}\label{Eq17}
        [\gamma^{0}p_{0}+\gamma^{1}p_{x}-\gamma^{0}V(x)-m(x)]\psi(x,t)=0.
    \end{align}

Now, let us use Majorana's representation, namely,
    \begin{align}\label{Eq18}
        \gamma^{0}=\sigma^{2}=\begin{pmatrix}
    0 & -i\\
    i & 0\\
    \end{pmatrix} \quad  \text{and} \quad   \gamma^1=i\sigma^{3}=\begin{pmatrix}
    i & 0\\
    0 & -i\\
    \end{pmatrix},
    \end{align}
    with $(\gamma^{0})^{2}=\mathbb{I}$. 
    
For this system, the Hamiltonian operator is
    \begin{align}\label{Eq19}
        \hat{H}=\alpha p_{x}+\beta m(x)+V(x),
    \end{align}
where $\alpha=\gamma^{0}\gamma^{1}$ and $\beta=\gamma^{0}$. 
    
To achieve our purpose, allow us to assume that the wave function $\psi(x, t)$ is
    \begin{align}\label{Eq20}
        \psi(x,t)=\begin{pmatrix}
            \varphi(x)\\
            \chi(x)\\
        \end{pmatrix}\text{e}^{-iEt}.
    \end{align}
This separation of variables leads us to the system of equations:
    \begin{align}\label{Eq21}
        i\bigg[\dfrac{d\chi(x)}{dx}-m(x)\chi(x)\bigg]=(E-V)\varphi(x),
    \end{align}
and
    \begin{align}\label{Eq22}
        i\bigg[\dfrac{d\varphi(x)}{dx}+m(x)\varphi(x)\bigg]=(E-V)\chi(x).
    \end{align}
    
Decoupling the Eqs. (\ref{Eq21}) and (\ref{Eq22}), one obtains
    \begin{align}
    \label{Eq23}
       -\varphi''(x)+\left[-m'(x)-2EV(x)+V^{2}(x)\right]\varphi(x)=[E^{2}-m^{2}(x)]\varphi(x),
    \end{align}
and
    \begin{align}
    \label{Eq24}
        \chi''(x)-\left[-m'(x)-2EV(x)+V^{2}(x)\right]\chi(x)=[E^{2}-m^{2}(x)]\chi(x).
    \end{align}
In this scenario, the prime notation denotes the derivative concerning the position variable. We are interested in studying the description of fermions with PDM, such as Majorana's particles with exciton-like PDM without impurities. Thus, one should consider $V(x)=0$. This consideration leads us to the Schr\"{o}dinger-like equation, i.e.,
    \begin{align}
    \label{Eq25}
        -\varphi''(x)+[m^{2}(x)-m'(x)]\varphi(x)=E^{2}\varphi(x).
    \end{align}
    
Comparing Eq. (\ref{Eq25}) with the usual Schrödinger theory, it is noted that even $V(x)=0$, the system will exhibit effective interaction\footnote{This interaction is called effective potential.} due to the mass distribution. In this case, the effective potential is
    \begin{align}
        \label{Eq26}
        V_{\text{eff}}=m^{2}(x)-m'(x).
    \end{align}
    
\begin{figure}[!ht]
        \centering
        \includegraphics[height=6cm,width=7cm]{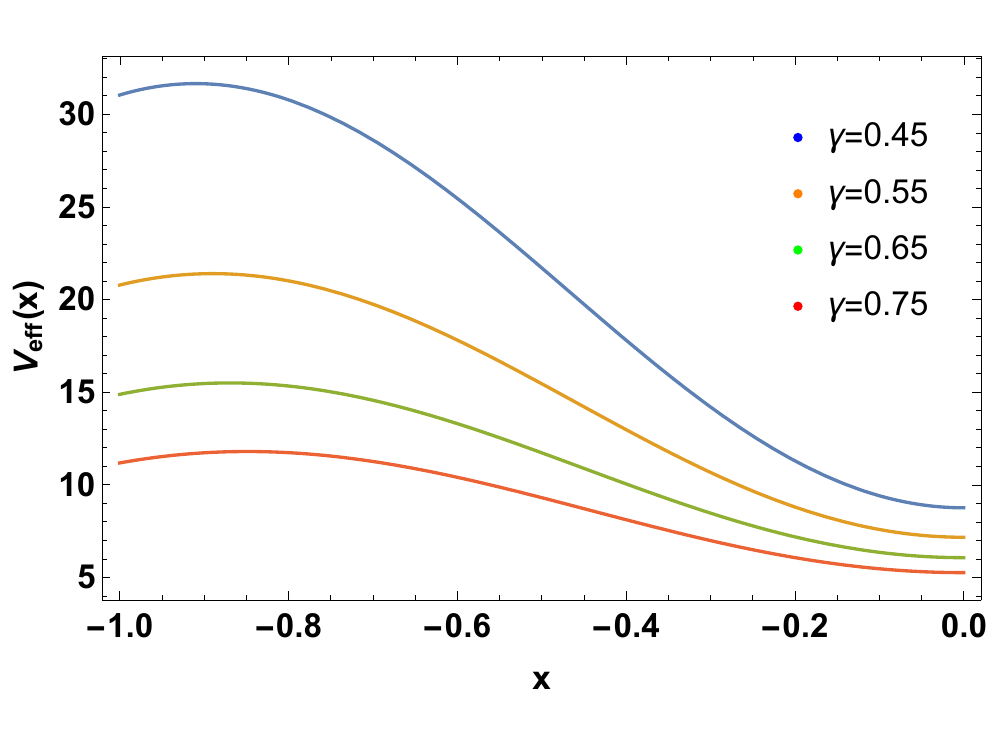}\vspace{-0.5cm}
        \caption{The effective potential $V_{\text{eff}}\,$ vs. $x$.}
        \label{fig2}
    \end{figure}

Replacing the definition of effective potential in Eq. \eqref{Eq26}, we conclude that
    \begin{align}\label{Eq27}
        V_{\text{eff}}=\dfrac{2\pi}{\gamma^{2}}\left[\sin^{2}{\left(\dfrac{\pi x}{2}\right)}+\dfrac{\gamma}{4}\sqrt{2\pi}\cos{\left(\dfrac{\pi x}{2}\right)}\right].
    \end{align}
This potential is half of a Pöschl-Teller-like potential \cite{Radovanovic,Sun,Miranda}. We expose this potential in Fig. \ref{fig2}.
    
Analyzing the effective potential (\ref{Eq27}), we find that Majorana's equation for the spinor $\varphi(x)$ with exciton-like PDM is
    \begin{align}
    \label{Eq28}
        -\varphi^{''}(x)+\left[\dfrac{2\pi}{\gamma^{2}}\sin^{2}{\left(\dfrac{\pi x}{2}\right)}+\dfrac{\pi\sqrt{2\pi}}{2\gamma}\cos{\left(\dfrac{\pi x}{2}\right)}\right]\varphi(x)=E^{2}\varphi(x),
    \end{align}
    where $x\in [-1,0]$.
    
\begin{figure}[!ht]
  \centering
  \subfigure[The case $\gamma=0.45$.]{\includegraphics[height=6cm,width=7cm]{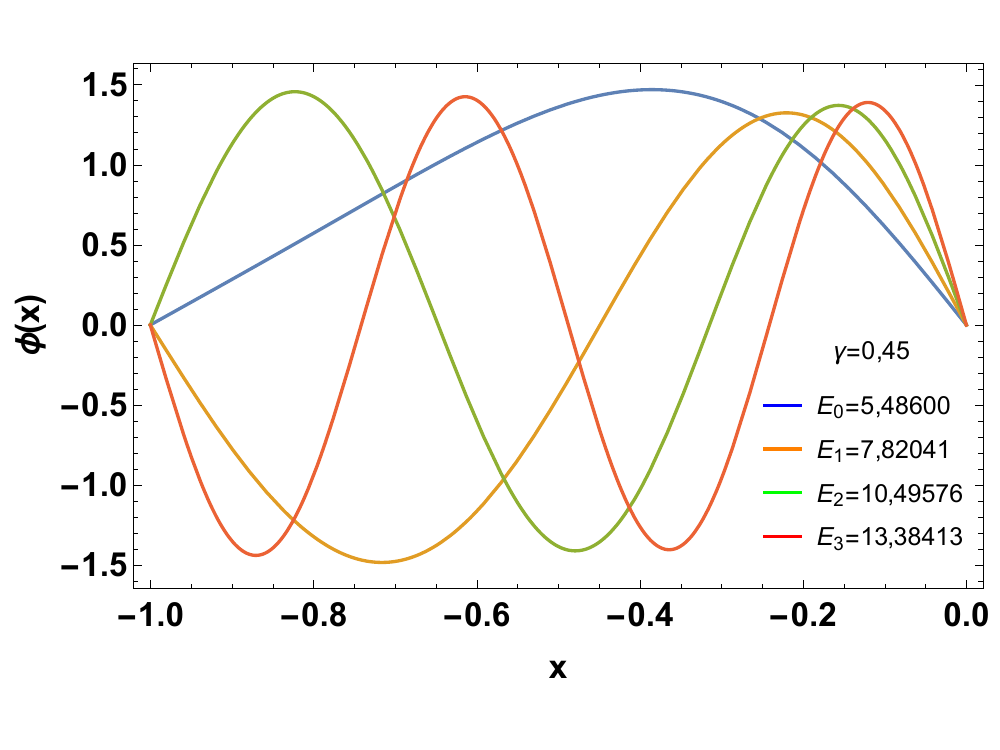}}
  \subfigure[The case $\gamma=0.55$.]{\includegraphics[height=6cm,width=7cm]{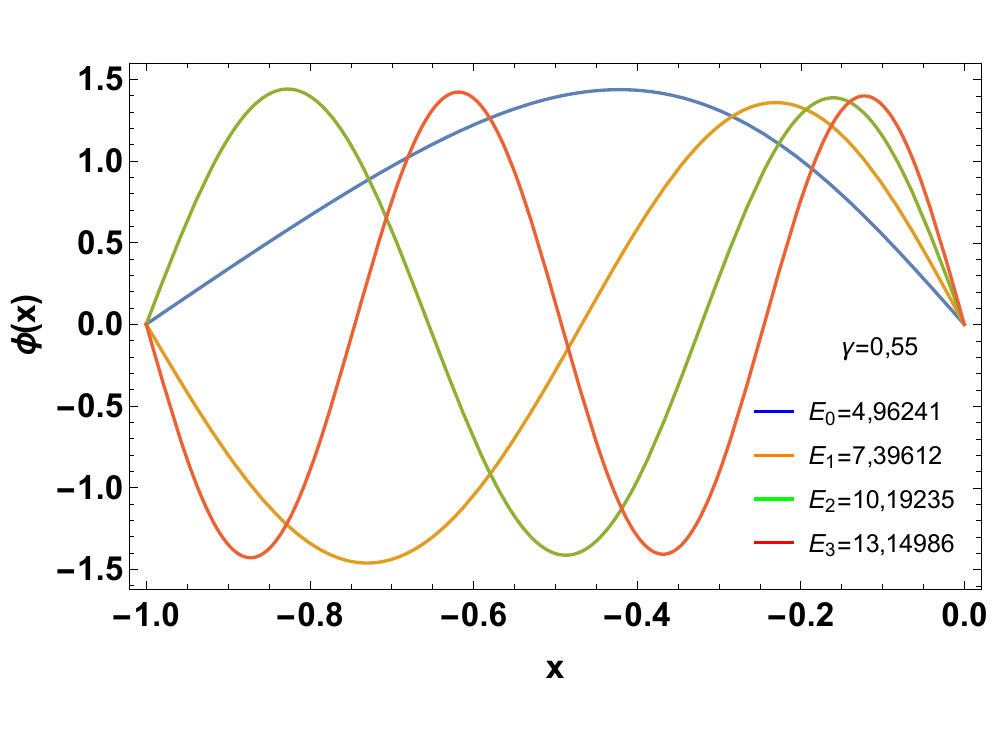}}\hfill
  \subfigure[The case $\gamma=0.65$.]{\includegraphics[height=6cm,width=7cm]{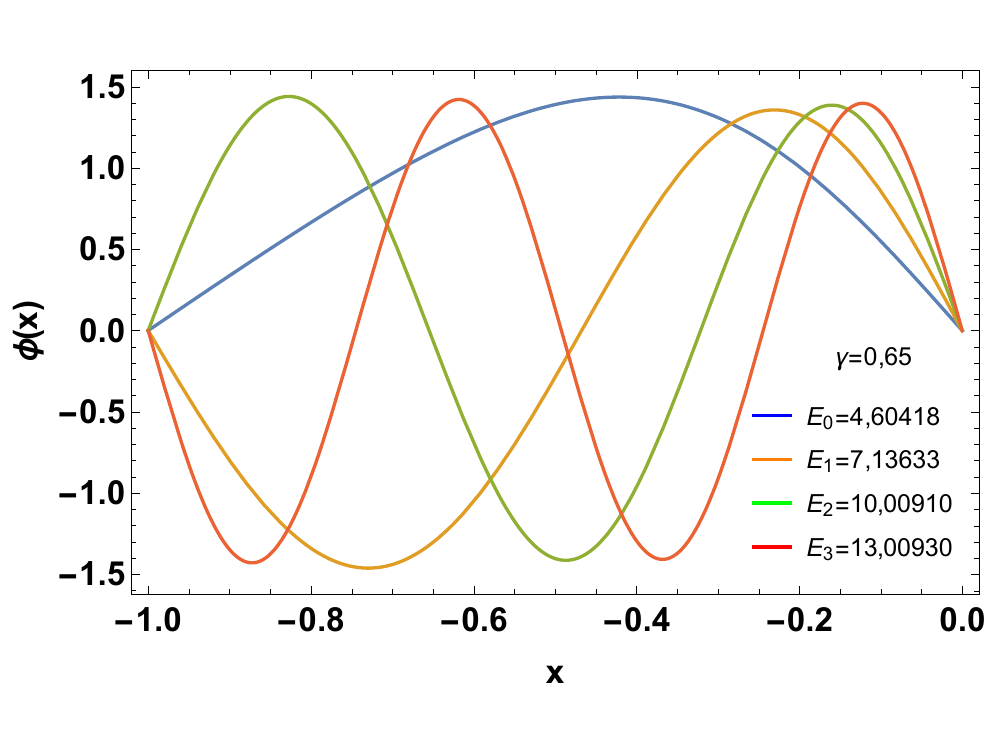}}
  \subfigure[The case $\gamma=0.75$.]{\includegraphics[height=6cm,width=7cm]{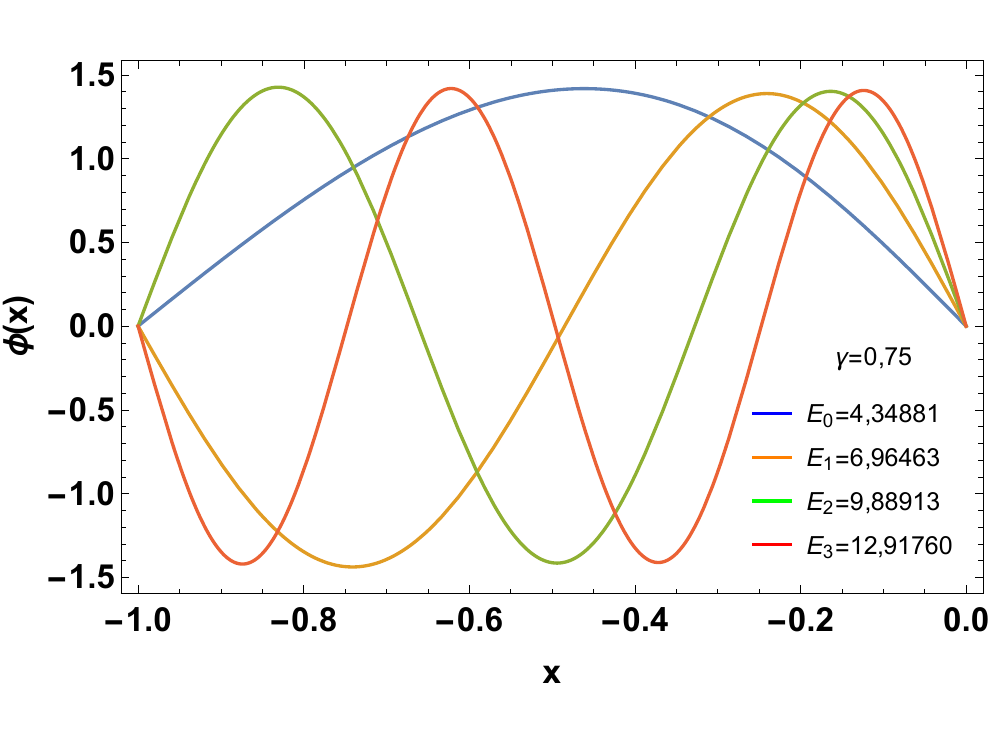}}\hfill
\caption{Wavefunctions $\varphi(x)$ vs. $x$ for the first four energy levels.}
        \label{fig3}
\end{figure}

Eq. (\ref{Eq28}) only admits numerical solutions in the interval $[-1,0]$\footnote{This range is due to the mass profile corresponding to the first Brillouin zone.}. Therefore, it is necessary to employ a numerical method to accurately describe the analytical solution of the wave function $\varphi(x)$. Thus, we utilize the finite element method\footnote{One uses the finite element method to find approximate solutions to differential equations \cite{Hildebrand}. This method involves discretizing the space corresponding to the interval $[-1,0]$ with steps of $10^{-2}$ \cite{Hildebrand}. Thus, we determine the solutions within each finite element in the range.}, adopting the Dirichlet condition\footnote{The Dirichlet boundary condition specifies the values which function must take on a specific part of the boundary in the domain where the differential equation admit solution \cite{Hildebrand}.}. Considering this approach, one obtains the numerical system's solutions. We expose these solutions in Fig. \ref{fig3}[(a)-(d)].

\begin{figure}[!ht]
    \centering
    \includegraphics[height=6cm,width=7cm]{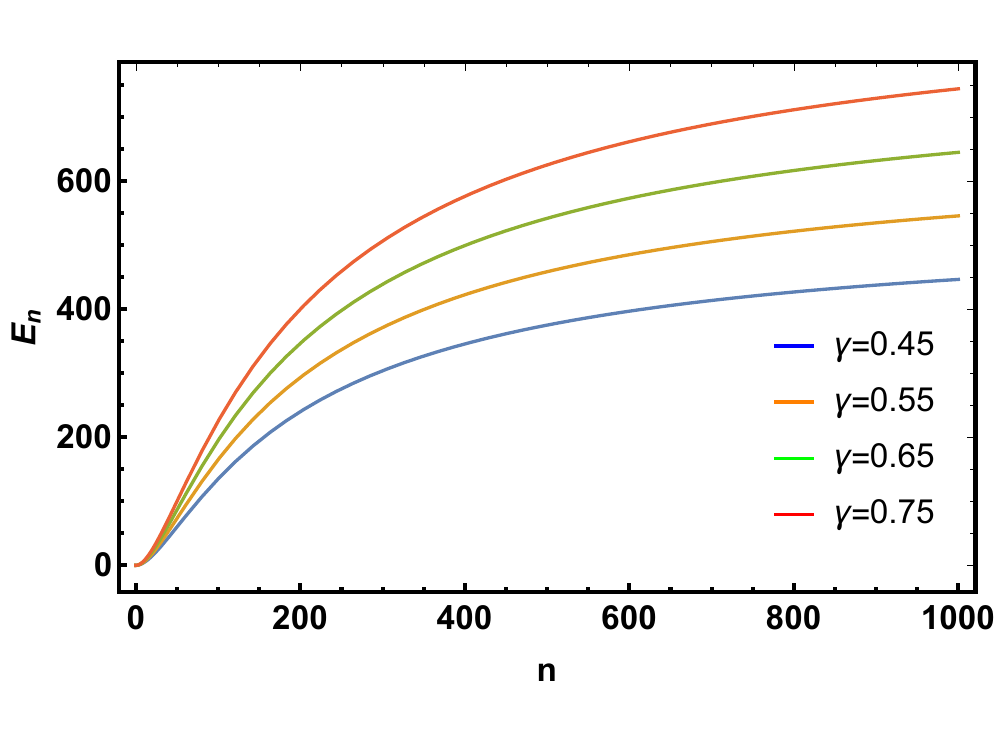} \vspace{-0.5cm}
    \caption{Numerical solutions of the energy eigenvalues ($E_n$) vs. quantum number $n$ for several values of the $\gamma$ parameter.}
    \label{fig4}
\end{figure}

By numerical inspection, we observe that the system admits bound states. That is an attractive result once all interactions in relativistic theory arise from mass distribution. Consequently, we can infer that Majorana's fermion with exciton-like PDM induces a self-interaction leading to quantizable eigenstates. One presents the detailed numerical quantization of the first energy vs. quantum number in Fig. \ref{fig4}.
    
\section{Thermodynamic properties}\justifying
    
From now on, we will study the thermodynamic properties of the system. To reach our purpose, let us assume, a priori, the construction of a Majorana fermion ensemble with exciton-like PDM. In this framework, the canonical partition function\footnote{In a canonical ensemble, the particle number remains constant, which leads us to a chemical potential null. That occurs because the particle number $N$ remains constant \cite{Pathria}.} \cite{Oliveira,Pathria} is 
    \begin{align}\label{Eq29}
        Z = \prod_{n} \left( 1 + \text{e}^{-\beta E_n} \right),
    \end{align}
where $\beta=1/k_B T$. Here, $k_B$ is Boltzmann's constant, and $T$ is the thermal bath temperature. 
    
Using Eq. (\ref{Eq29}), we can obtain all the thermodynamic properties of a Majorana fermion ensemble with exciton-like PDM in a thermal bath. These properties are Helmholtz free energy, internal energy, entropy, and heat capacity. These quantities are, respectively, 
\begin{align}\label{Eq30}
    \mathcal{F}=-\frac{1}{\beta}\text{ln} Z_N, \hspace{0.5cm}
    U=-\frac{\partial}{\partial \beta}\text{ln} Z_N, \hspace{0.5cm}
    S=k_B\beta^2 \frac{\partial F}{\partial \beta}, \hspace{0.5cm} \text{and}\hspace{0.5cm} 
    C_v=-k_B\beta^2\frac{\partial U}{\partial \beta}.
\end{align}
Here, $Z_N$ is the total partition function for the ensemble of indistinguishable N-fermions in a thermal bath, 

Note that we obtain the eigenstate results numerically. Thus, we must adopt a numerical approach to access the thermodynamic properties. In this form, in the subsection below, one presents the numerical results of the thermodynamic properties.

\subsection{Numerical results}\justifying

\begin{figure}[!ht]
  \centering
  \subfigure[The Helmholtz free energy.]{\includegraphics[height=6cm,width=7cm]{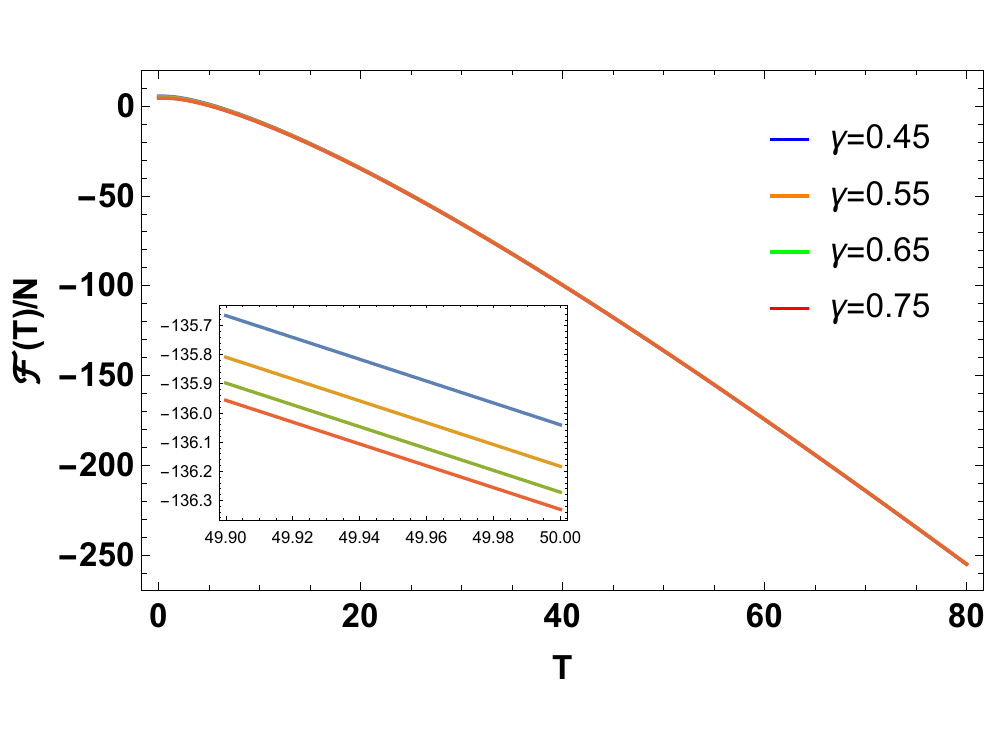}}
  \subfigure[The mean energy.]{\includegraphics[height=6cm,width=7cm]{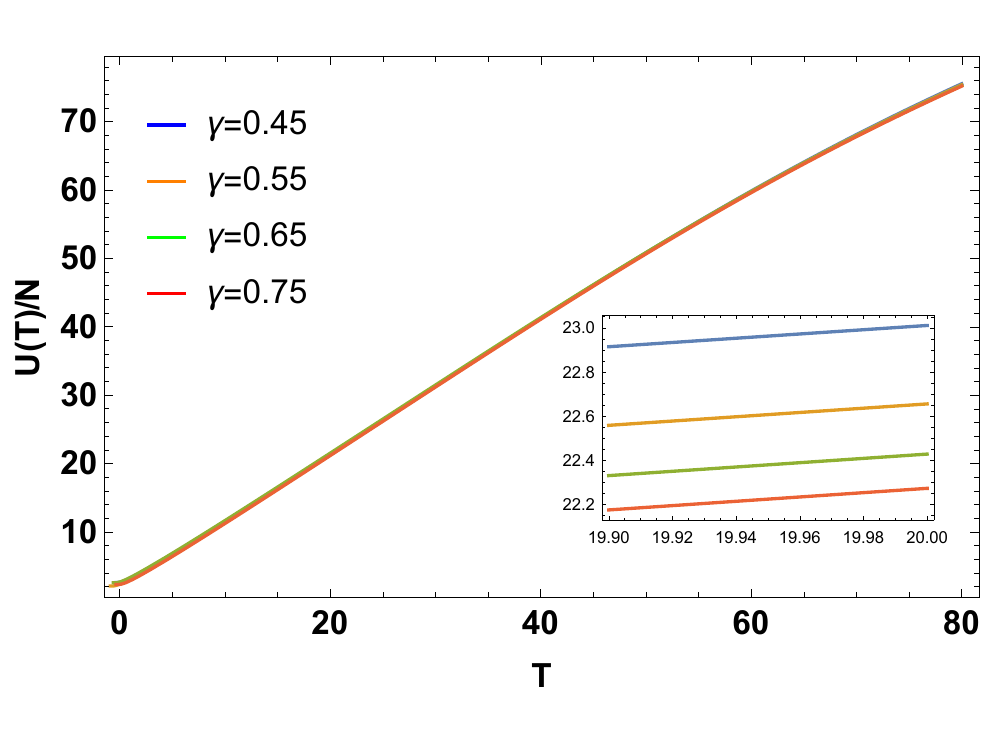}}\hfill
  \subfigure[The entropy.]{\includegraphics[height=6cm,width=7cm]{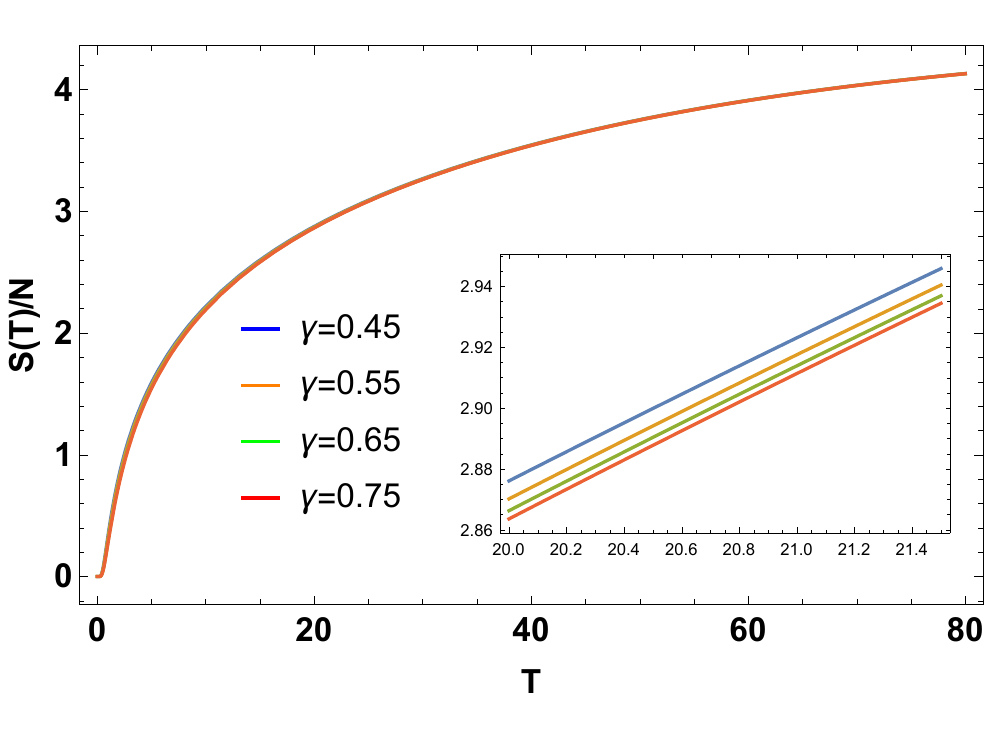}}
  \subfigure[The heat capacity.]{\includegraphics[height=6cm,width=7cm]{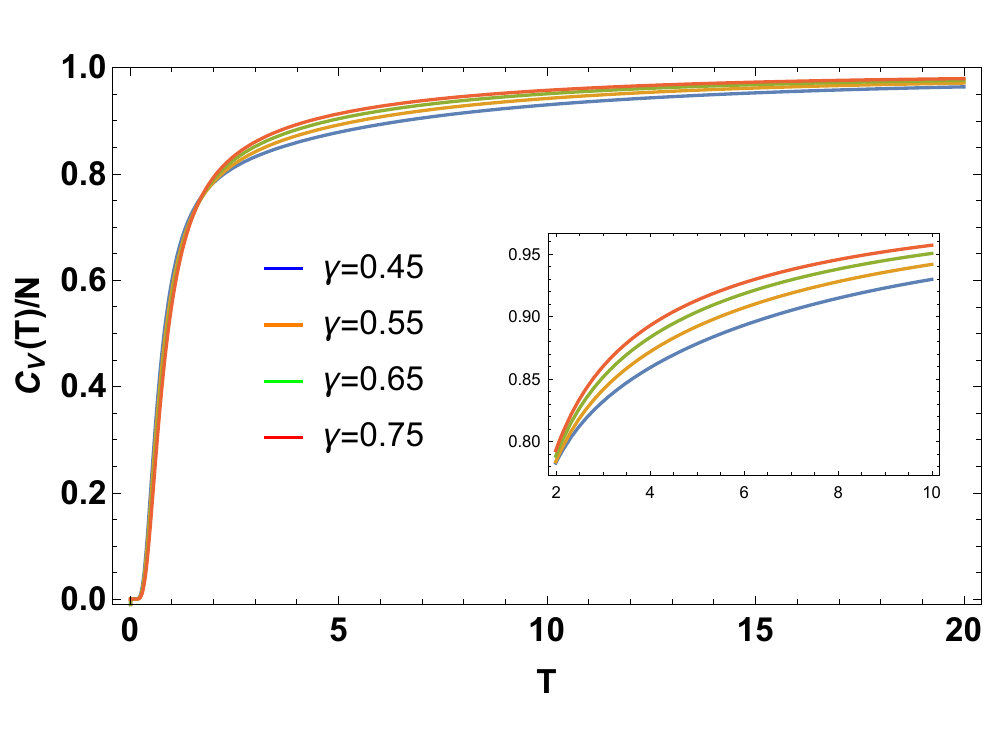}}\hfill
\caption{Thermodynamic properties of Majorana fermions with exciton-like PDM.}
        \label{fig5}
\end{figure}

Investigating the thermodynamic properties, we adopted the numerical solution shown in Fig. \ref{fig4}. Using the numerical solution for the energy spectrum, i.e., the numerical expression for $E_n$ [Figure \ref{fig4}], we explored the partition function associated with the Majorana fermion ensemble with PDM using the numerical interpolation method with steps of $10^{-2}$. Thus, we elucidated the behavior of the partition function (\ref{Eq29}). Using the numerical solution of the partition function, together with definitions of thermodynamic properties, we obtain results as displayed in Eq. (\ref{Eq30}). In this way, one investigates numerically the thermodynamic properties. In Fig. \ref{fig5} are the results of the thermodynamic properties.

In Figs. \ref{fig5}(a), \ref{fig5}(b), \ref{fig5}(c), and \ref{fig5}(d), we numerically plotted all profiles of thermal quantities as functions of temperature $T$ for different lattice parameter, i.e., $\gamma$. For one-dimensional Majorana fermions with exciton-like PDM, one notes that Helmholtz's function $\mathcal{F}(T)/N$ decreases for high-temperature values and exhibits higher values when the lattice parameter is small. In contrast, the mean energy $U(T)/N$ shows an almost linear behavior. As the lattice parameter increases, the curve of the mean energy undergoes a slight decrease. Meanwhile, the entropy $S(T)/N$ increases with $T$ and decreases as the lattice parameter increases. The heat capacity $C_V (T)/N$ tends toward an asymptotic behavior fixed at the unit value when $T$ increases.

From the analysis of the profiles of the curves of the aforementioned thermodynamic functions, we extract information about the thermal behavior of the one-dimensional Majorana fermion ensemble with exciton-like PDM. Physically, it is noticeable that Helmholtz's free energy has a critical point for low temperatures, i.e., $T\simeq 0$. Thus, one notes that the energy available for performing work decreases rapidly when the temperature increases. On the other hand, we note that when this ensemble tends to thermal equilibrium in the reservoir, the system's mean energy continuously increases. Furthermore, the change in the lattice's parameter tends to slow the thermodynamic behavior until the system reaches equilibrium. Finally, we observe that regardless of the lattice parameter, all quantities assume higher values at low temperatures, as predicted in the literature. Lastly, one notes that the well-known Dulong-Petit law is satisfied once that $C_{v}(T\to 0)/N\simeq k_B$.

\section{Summary and conclusion}\justifying

We investigated Majorana's excitons-like quasiparticles by using the PDM approach. In this context, we adopted the PDM concept to reach the purpose once this approach induces a theory for these quasiparticles. Furthermore, we have assumed that the quantum mechanical system is without impurities. In this framework, one adopts Majorana's fermions with exciton-like mass distribution to build and study the quantum dynamics of the theory. Besides, we employ statistical mechanics arguments to obtain the Helmholtz free energy, mean energy, entropy, and specific heat.

Assuming the quantum theory, we investigate Majorana's fermions with an exciton-like PDM. In this framework, one notes that the mass distribution induces an effective interaction purely due to the particle's mass, i.e., $V_{\text{eff}}=m^2(x)-m'(x)$. That potential at the first Brillouin zone results in half of a P\"{o}shl-Teller-like self-interaction, confining Majorana fermions and generating bound states resembling to the P\"{o}shl-Teller eigenstates. In other words, even in the absence of electromagnetic interaction, i.e., $ A_\mu= (0,0)$, the system will exhibit the existence of bound states purely due to the presence of the mass distribution. Thus, if one-dimensional Majorana fermions possess bound states with exciton-like mass, it admits bound states even in the case of free particles.

Furthermore, numerical results demonstrate that when the lattice parameter $\gamma$ is small, regions of maximum probability are near the boundary from the first Brillouin zone. Meanwhile, as the atoms become more spaced in the crystal lattice (i.e., as $\gamma$ increases), the probability regions exhibit a symmetrical probability distribution in the first Brillouin zone.

Considering a Majorana particle ensemble with exciton-like PDM in a thermal bath, we noted that the Helmholtz free energy has a maximum capacity to perform work as $T\to 0$ and decreases as the ensemble temperature increases. Thus, one concludes that the amount of free energy available for work decreases rapidly. Moreover, this available free energy is higher when the lattice parameter is smaller. Meantime, the mean energy continuously increases and is always higher for smaller $\gamma$ values. Additionally, the thermal variation of the particle ensemble tends to increase until reaching thermal equilibrium in the considered system. Indeed, this behavior resonates in the heat capacity, where reaching thermodynamic equilibrium remains constant, following the Dulong-Petit law. Finally, one can note that the system's entropy adheres to the basic principles of thermodynamics with a monotonically increasing behavior independent of the lattice parameter. However, the system's entropy increases rapidly for small $\gamma$ values. In this regime, the thermodynamic system will have more available energy.

Finally, one highlights that Majorana's particles possess topological properties once the particle and its antiparticle are identical. Naturally, we noted that these properties arise in the thermodynamic properties, which reveal a phase transition at a critical temperature, approximately $T \approx 1.25$. Below this temperature, the system may be in a topological phase, with the Majorana particles well-defined and Majorana's excitations announced. Above this temperature, Majorana excitations disappear, indicating thermodynamic stability. Furthermore, this topological nature suggests at least the existence of a zero-energy eigenstate. However, the exciton-like PDM violates this topological property due to the Pöschl-Teller-like effective potential.

\section{Acknowledgment}\justifying

The authors are grateful to the Conselho Nacional de Desenvolvimento Científico e Tecnológico (CNPq). F. C. E. Lima and C. A. S. Almeida are supported, respectively, for grants No. 171048/2023-7 (CNPq/PDJ) and 309553/2021-0 (CNPq/PQ).

\section{Conflicts of Interest/Competing Interest}\justifying

All the authors declared that there is no conflict of interest in this manuscript.

\section*{Appendix A. Determining $m(x)$}\justifying

Considering the expression for the dispersion energy presented in Eq. \eqref{Eq13}, i.e.,
\begin{align}
    \label{EqA1}
    m(k)=\frac{\hbar}{2\gamma a^2}\sec(k a).
\end{align}

To determine the mass distribution at the position space, i.e., $m(x)$, the Fourier transform is applied to Eq. \eqref{EqA1}, resulting in
\begin{align}
    \label{EqA2}
    m(x)=& \frac{\hbar}{2\gamma a^2\sqrt{2\pi}}\,\int_{-\frac{\pi}{a}}^{+\frac{\pi}{a}}\, \sec(k a)\cdot \text{e}^{-ikx}\,dk.
\end{align}

By applying the residue theorem to evaluate the integral in Eq. \eqref{EqA3} \cite{Nikiforov}, one obtains
\begin{align}
    \label{EqA3}
    m(x)=& \frac{\hbar}{2\gamma a^2 \sqrt{2\pi}}\cdot 2\pi i\cdot \left[\text{Res}\left(-\frac{\pi}{2a}\right)+\text{Res}\left(\frac{\pi}{2a}\right)\right],
\end{align}
where $\text{Res}\left(\pm\frac{\pi}{2a}\right)=\pm a^{-1}\text{e}^{\pm\frac{i\pi x}{2a}}$ are the residues concerning the respective poles $k=\pm\frac{\pi}{2a}$, which leads us to
\begin{align}
    \label{EqA4} \nonumber
    m(x)=& \frac{\hbar}{2\gamma a^2 \sqrt{2\pi}}\cdot 2\pi i\cdot \left[\text{Res}(-\pi/a)+\text{Res}(\pi/a)\right]\\  \nonumber
    m(x)=& \frac{\hbar}{2\gamma a^2 \sqrt{2\pi}}\cdot 2\pi i\cdot \left[\frac{\text{e}^{\frac{i\pi x}{2a}}}{a}-\frac{\text{e}^{-\frac{i\pi x}{2a}}}{a}\right]\\  \nonumber
    m(x)=& \frac{i^2\hbar\sqrt{2\pi}}{\gamma a^3}\left(\frac{\text{e}^{\frac{i\pi x}{2a}}-\text{e}^{-\frac{i\pi x}{2a}}}{2i}\right)\\  
    m(x)=& -\frac{\hbar\sqrt{2\pi}}{\gamma a^3}\sin\left(\frac{\pi x}{2a}\right).
\end{align}
Therefore, by assuming natural units ($\hbar=1$) and the unit lattice parameter ($a=1$), one obtains
\begin{align}
    \label{EqA5}
    m(x)=-\frac{\sqrt{2\pi}}{\gamma}\sin\left(\frac{\pi x}{2}\right).
\end{align}

\end{document}